\documentclass[letter]{aa}  
\usepackage{graphicx}
\usepackage{hyperref}                                                     
\usepackage{color}
\usepackage{txfonts}
\newcommand{\kms}{km\,s$^{-1}$}

\newcommand{\Msun}{\,\mathrm{M}_\odot}

\begin{document}

   \title{SO as shock tracer in protoplanetary disks: the AB Aurigae case}

   \author{A. Dutrey  \inst{1}
          \and E. Chapillon \inst{1,2}   
          \and S. Guilloteau \inst{1}
          \and Y.W. Tang \inst{3}
          \and A.Boccaletti \inst{4}
          \and L.Bouscasse \inst{2}
          \and Th. Collin-Dufresne \inst{1}
          \and E. Di Folco \inst{1} 
          \and A. Fuente \inst{5}
          \and V.Pi\'etu \inst{2}
          \and P. Rivi\`ere-Marichalar \inst{5}
          \and D.Semenov \inst{6,7}
          }

   \institute{Laboratoire d'Astrophysique de Bordeaux, Universit\'e de Bordeaux, CNRS, B18N, 
   All\'ee Geoffroy Saint-Hilaire, F-33615 Pessac\\
              \email{anne.dutrey@u-bordeaux.fr}
    \and IRAM, 300 Rue de la Piscine, F-38046 Saint Martin d'H\`{e}res, France
    \and Academia Sinica Institute of Astronomy and Astrophysics, PO Box 23-141, Taipei 106, Taiwan
    \and LESIA, Observatoire de Paris, Université PSL, CNRS, Sorbonne
Université, Univ. Paris Diderot, Sorbonne Paris Cité, 5 place Jules
Janssen, 92195 Meudon, France
   \and Centro de Astrobiología (CAB), CSIC-INTA, Ctra Ajalvir Km 4, Torrejón de Ardoz, 28850 Madrid, Spain
   \and Max-Planck-Institut f\"{u}r Astronomie (MPIA), K\"{o}nigstuhl 17, D-69117 Heidelberg, Germany
\and Department of Chemistry, Ludwig-Maximilians-Universit\"{a}t, Butenandtstr. 5-13, D-81377 M\"{u}nchen, Germany}
   \date{\today}

 
  \abstract
   {Sulfur Monoxide is known to be a good shock tracer in molecular clouds and protostar environments, but its abundance is difficult to reproduce even using state-of-the-art astrochemical models. }
   {We investigate the properties of the observed SO emission in the protoplanetary disk of AB Auriga, a Herbig Ae star of 2.4 $\Msun$ located at 156 pc. The AB Aur system is unique because it exhibits a dust trap and at least one young putative planet orbiting at about 30 au from the central star.}
   {We reduced ALMA archival data (projects 2019.1.00579.S, 2021.1.00690.S and 2021.1.01216.S) and analyzed the three detected SO lines (SO $6_5-5_4$, $6_7-5_6$ and $5_6-4_5$). 
   We also present C$^{17}$O and C$^{18}$O 2-1 data to complement the interpretation of the SO data.}
   {For the three SO lines, the maximum SO emission in the ring is not located in the dust trap. Moreover, the inner radius of the SO ring is significantly larger than the CO emission inner radius, $\sim 160$ au versus $\sim 90$ au. The SO emission traces gas located in part beyond the dust ring. This emission likely originates from shocks at the interface of the outer spirals, observed in CO and scattered light emission, and the molecular and dust ring. SO is also detected inside the cavity, at a radius $\sim 20-30$ au and with a rotation velocity compatible with the proto-planet P1.  We speculate that this SO emission originates from accretion shocks onto the circumplanetary disk of the putative proto-planet P1.}
  {These observations confirm that SO is a good tracer of shocks in protoplanetary disks and could be a powerful, new tool to detect embedded (proto-)planets.}


   \maketitle
%
\nolinenumbers

\section{Introduction}

While exoplanet detections are mostly done at optical and near-infrared (NIR) wavelengths toward disks where dense gas has been dissipated, the observation of young embedded protoplanets, for which mass accretion is still ongoing, is fundamental to constrain models of planet formation. 

Large facilities such as ALMA or SPHERE/VLT have imaged proto-planetary disks showing complex gas and dust structures, such as large central cavities, gaps, asymmetries, and spiral patterns (e.g. HD 142527 \cite{Benisty+2015} and \cite{Fukagawa+2006};  MWC758 \cite{Benisty+2015}, \cite{Grady+2013} and \cite{Isella+2010}; HD135344B \cite{Garufi+2013} and \cite{Muto+2012ApJ}; Elias 2-27 \cite{Perez+2016}, and AB Aurigae \cite{Fukagawa+2004}, \cite{Tang+2012} and \cite{Tang+2017}). ALMA has also revealed velocity kinks in the Keplerian profile \citep[e.g.][and references therein]{Pinte+2018}.
These features, which can be reasonably regular, asymmetric, or more reminiscent of faint discontinuous disks, are due to gravitational disturbances, other (magneto-)hydrodynamical processes, and dust pressure traps.  
In several cases, they indirectly reveal disturbers, such as embedded protoplanets (e.g., \cite{Zhu+2015}), which cannot be directly observed in gas-rich disks. So far, young planets have been detected through direct imaging of the associated circumplanetary disk only in the case of PDS 70 \citep{Keppler+2018,Isella+2019}.

\begin{figure}
\includegraphics[width=9cm]{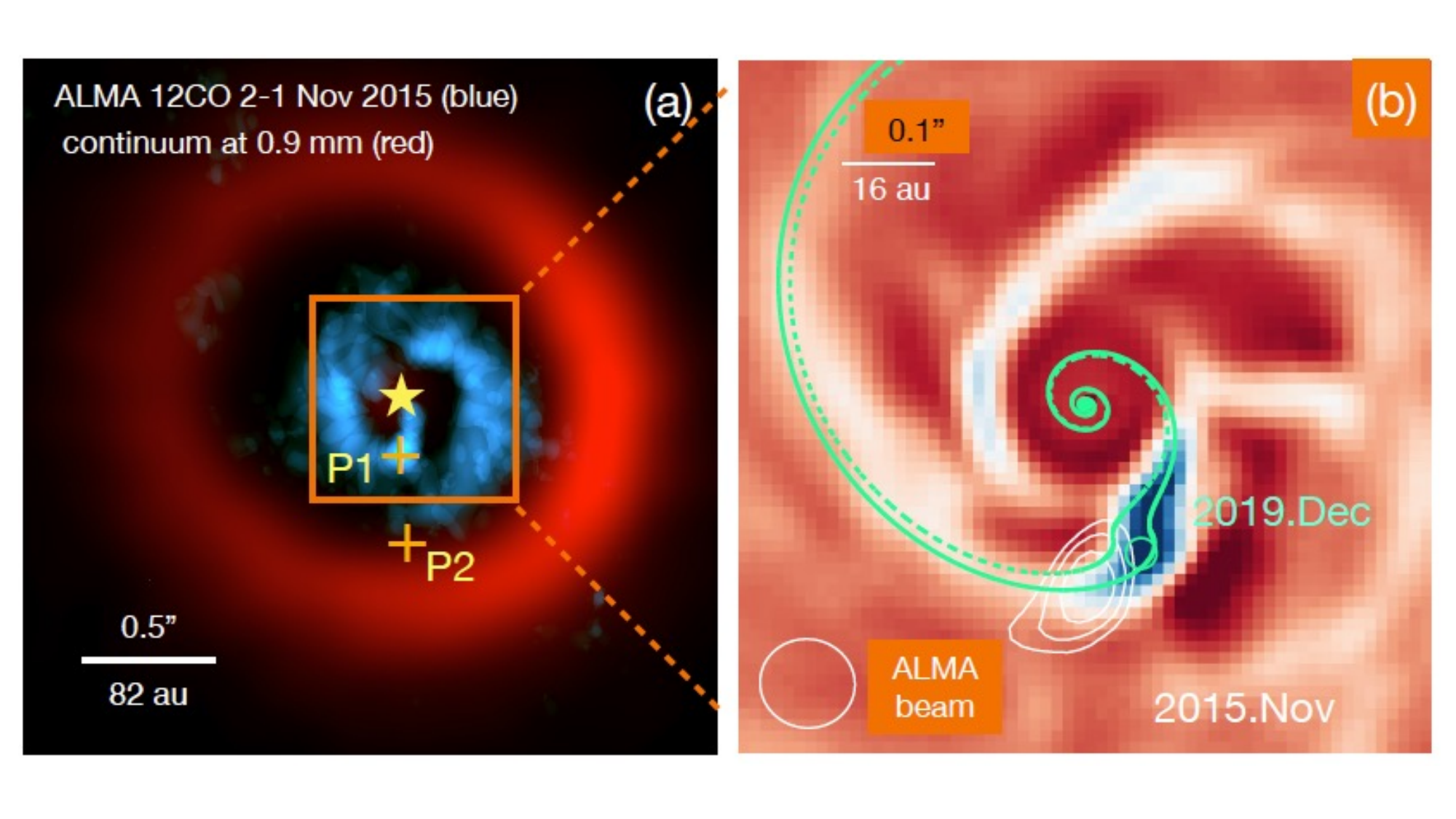}
\caption{Spirals and clump-like protoplanets in AB Aur transition disk. (a) ALMA image \citep{Tang+2017} of the $^{12}$CO J=2-1 (cyan) inner spirals detected inside the mm dust ring (red) with the two putative planets (P1 and P2 crosses). (b) Spirals with a twist structure detected in NIR using SPHERE/VLT \citep{Boccaletti+2020}. The green solid curve marks the best-fit model reproducing the spiral twist near the perturber (P1, green circle). The CO spot is in white contours. The green dashed curve marks the model offset by 14.1$^o$, corresponding to the Keplerian rotation in 4 yrs between ALMA (nov.2015)  and SPHERE (dec.2019) data. The two bright spots (CO and NIR) are separated by a distance in agreement with an embedded planet rotating at the Keplerian speed measured from previous CO observations. The white arrow marks the rotation direction of the disk.}
\label{Fig:ABAUR}%
\end{figure}
A key parameter to determine young planet properties is the accretion rate. Its measurement can be performed using the H$_\alpha$ line \citep{Haffert+2019,Zhou+2021,Zhou+2022}, even if this has been recently debated \citep{Zhou+2023}. Identifying additional observable tracers of planets in disks will allow us to better detect and characterise them. At mm wavelengths, SO is known to be an efficient tracer of shocks in various environments such as the interstellar medium \citep{Prasad+1980} and around young stars \citep[e.g.][]{Sakai+2014,VanGelder+2021,Garufi+2022}. Using the 30-m radio telescope, \cite{Fuente+2010} 
reported the first detection of SO in a proto-planetary disk, the disk around AB Aurigae, which was then imaged using NOEMA by \citet{Pacheco+2016}. 
More recently, SO has also been detected in the dust trap around Oph-IRS 48 \citep{Booth+2021} and found in proto-planetary disks where it is suspected to trace embedded proto-planets \citep{Booth+2023,Law+2023}. 

Based on ALMA archival data, we investigate the behaviour of SO in the proto-planetary system of AB Aurigae (hereafter AB Aur). After describing the complex AB Aur environment, we present the observations and the results of our SO modelling. We then discuss the SO properties in this young planetary system. 

\section{Young planetary system around AB Aurigae }

Located at 156 pc \citep{Gaia+2022} and surrounded by a large gas-rich disk with a central cavity \citep{Pietu+2005}, AB Aurigae is one of the closest Herbig Ae stars (with a spectral type A0-A1 and a mass 2.4 $\Msun$). Its surroundings have been intensively studied from the optical (e.g. HST-STIS imaging \cite{Grady+1999}) up to the mm wavelengths \citep{Corder+2005}. Recent ALMA \citep{Tang+2017} and SPHERE observations \citep{Boccaletti+2020} show that the disk consists of three parts. There is an inner dust and CO disk of outer radius $\sim$ 20 au. Beyond this, a large cavity is observed both in CO gas and in the continuum, extending up to about 90-100 au \citep{Pietu+2005,Tang+2012}. Finally, an outer gas and dust disk (or ring) extends to about 200 au in continuum and more than 1000 au in CO \citep{Tang+2012,Pietu+2005}. Recent ALMA continuum images reveal that the dust ring appears asymmetric, 
the western part being much brighter. This enhancement is interpreted as a large dust trap, possibly associated with a decaying vortex \citep{Fuente+2017}, and will be referred to as ``the dust trap'' thereafter.

Thanks to a favourable low inclination ($i~\sim$22$^\circ$), the ring and its central cavity are well resolved, allowing images to reveal that the system exhibits two sets of spirals of different physical origins. 1) The outer spirals \citep{Fukagawa+2004,Tang+2012}) are detected at a large distance from the star ($> 2''$), and they result from the accretion of non-planar outer material onto the ring \citep{Tang+2012}. 2) In the cavity, the inner spirals, both detected in CO and NIR, are morphologically well explained by the formation of one or two massive planets \citep{Tang+2017,Boccaletti+2020,Currie+2022}). Figure \ref{Fig:ABAUR}, a montage from several papers, summarizes the main known properties of the inner part of this complex system. 
The first object (location P1 in Fig.\ref{Fig:ABAUR}), seen as a bright CO spot in \cite{Tang+2017}, is at the radius of $\sim$ 30 au, as expected from models \citep{Dong+2015}. Associated with P1, the polarized intensity image from SPHERE further reveals a twist in the NIR emission distribution
\citep{Boccaletti+2020}. Taking into account the time baseline between the CO and NIR observations (4 years) and the measured Keplerian speed, the locations of the bright CO spot and NIR twist coincide with the astrometric accuracy, as it would be the case for an embedded planet orbiting at r$\sim$30 au. In addition, the CO spot at P1 appears in several velocity channels, as predicted by models \citep{Perez+2015}. 

The second object (often called AB Aur b), at r$>$80 au (location P2 in Fig.\ref{Fig:ABAUR}) was reported by \cite{Currie+2022} from NIR observations with Subaru/CHARIS and the H$\alpha$ accretion tracer 
 \citep{Zhou+2022}. Recent UV and optical data seriously questioned the accretion scenario onto a young planet, \citep{Zhou+2023}. However, such an additional planet at P2, may explain the existence of the western spiral \citep{Tang+2017,Dong+2015} and the dust trap \citep{Baruteau+2021}. 

\section{Observations and Results}

\subsection{Observations}

To investigate the SO molecular distribution in the AB Aur system, we used ALMA archival data from projects 2019.1.00579 (PI Fuente), 2021.1.00690  (PI Dong) and 2021.1.01216.S (PI Huang). The calibrated measurement sets were provided by the ESO ARCnode. We then exported the visibility data as UVFITS format files, using the \textsc{CASA} task \texttt{exportuvfits}. These were transferred to the \textsc{GILDAS} data format for further processing using the \textsc{Imager}\footnote{See https:://imager.oasu.u-bordeaux.fr} program.
Data were corrected for proper motions using coordinates from GAIA (ICRF, $RA$=\,04:55:45.846 and $Dec$=\,+30:33:04.292,  $\mu\mathrm{RA} = 4.018 \pm 0.039$ and $\mu\mathrm{Dec} = -24.027 \pm 0.028$ mas/yr, \cite{Gaia+2022}). Self-calibration solutions were derived from the continuum data and applied to all spectral windows before performing imaging. The observed spectral lines used in the analysis are given in Table \ref{tab:obs} with the corresponding angular resolution and integrated fluxes. The spectral resolution of the data cubes ranges from 0.15 to 0.17 km\,s$^{-1}$. The continuum properties are also presented. Note that C$^{17}$O and C$^{18}$O 2-1 observations are only used to better characterise the SO emission properties and will be presented more specifically in a separate forthcoming paper.  

\begin{table}
\begin{tabular}{c|c|lr|lr}
\hline
 Line   & Frequency & Resolution  & PA  & Int. Flux\\
 & (MHz) & ($'' \times ''$) & ($^\circ$) & (Jy.km/s) \\
  \hline
C$^{17}$O 2-1   & 224714.385 &   $0.63 \times 0.34$ & 25 & 2.96 $\pm$ 0.64\\
C$^{18}$O 2-1   & 219560.354 &   $0.67 \times 0.42$ &  179 & 5.34 $\pm$ 0.02\\
SO $5_6-4_5$   &  219949.442 &  $0.44 \times  0.32$ & 170 & 0.33 $\pm$ 0.01\\
SO $6_5-5_4$   &  251825.770 & $ 0.51 \times 0.31$ & 167 & 0.27 $\pm$ 0.01\\
SO $6_7-5_6$   &  261843.721 & $ 0.49 \times 0.29$ & 168 & 0.27 $\pm$ 0.02 \\
\hline 
\hline
Cont. & (MHz) & ($'' \times ''$) & ($^\circ$) & (mJy) \\ 
\hline
  & 249000  & $0.42 \times 0.25$ & 168 & $82.7 \pm 0.4$ \\
\hline
\end{tabular}
\caption{Observed frequencies, transitions and continuum.}
\label{tab:obs}
\end{table}
\begin{figure}
 \centering
\includegraphics[width=0.8\columnwidth]{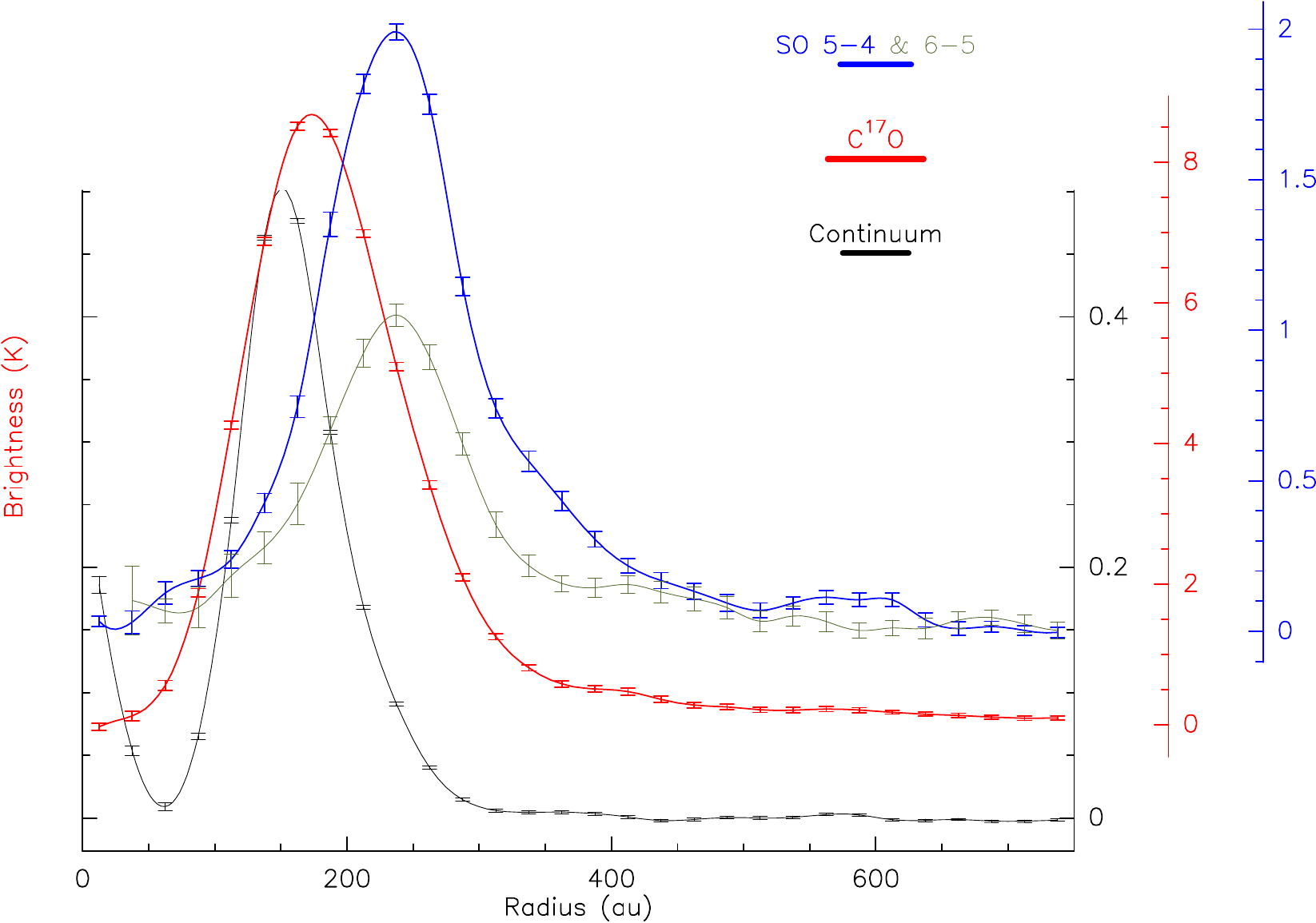}
\caption{Azimuthally averaged radial profiles of SO 5$_6$-4$_5$ and average
of the 6$_5$-5$_4$ and 6$_7$-5$_6$ lines,  C$^{17}$O J=2-1 line and the continuum emission at 1.3mm. The error bars indicate the linear resolution for each data set. The respective temperature scales are given on the right axes. The molecular radial profiles are derived from the Keplerian correction (see Appendix A).}
\label{fig:radial}%
\end{figure}
\begin{figure}
 \centering
\includegraphics[width=0.8\columnwidth]{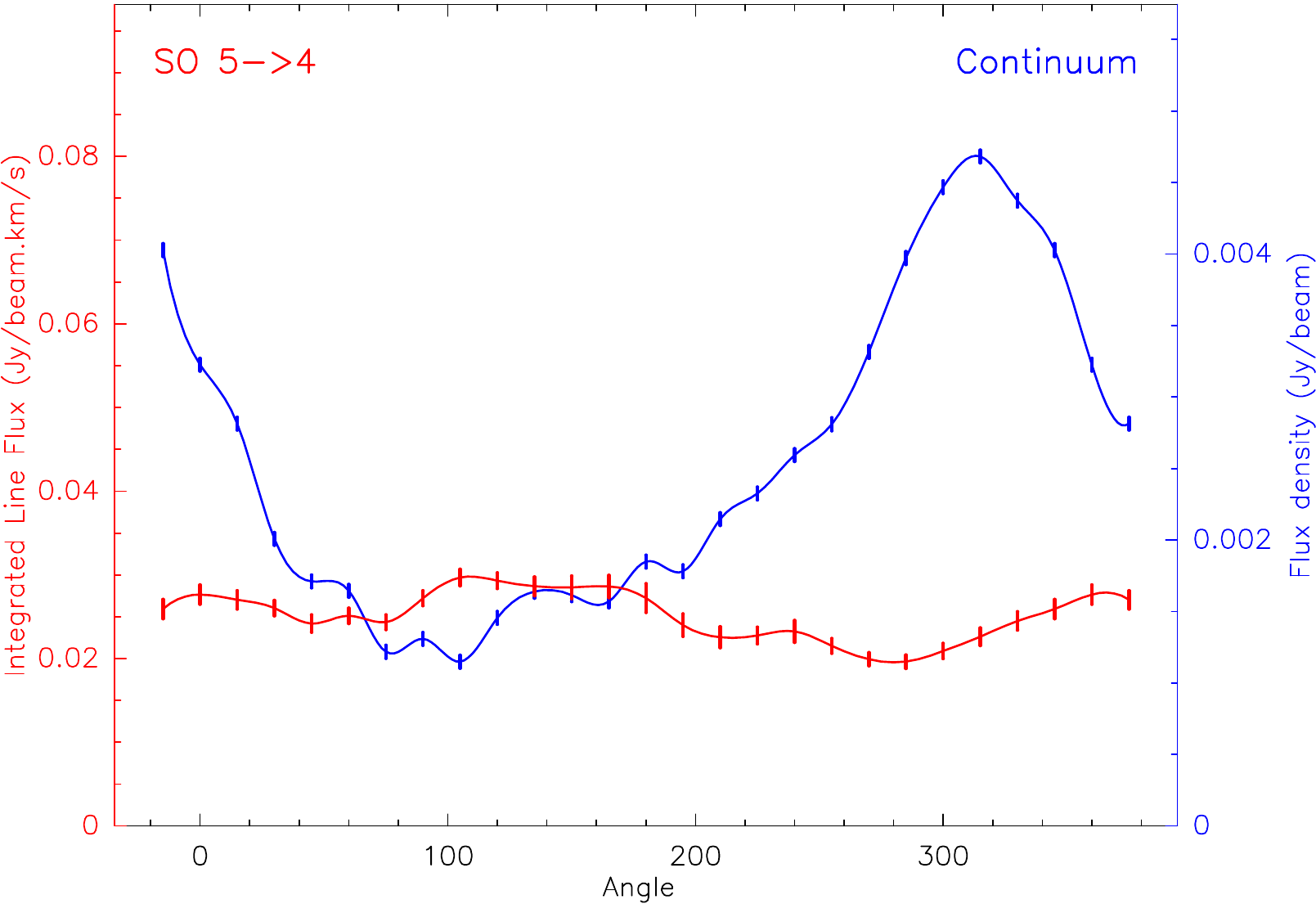}
\caption{Azimuthal profiles of the continuum in (blue) and SO\,$5_6-4_5$ (red)
emission, near their peak emission radii, 145 au and 238 au, respectively (see Fig.\ref{fig:radial}). The amplitude scales span 5 times the minimum value
for each profile.
\label{fig:azimut}%
}
\end{figure}
\subsection{Results}

\paragraph{Images and profiles}

Figure \ref{fig:areas} shows integrated area maps compared to the continuum image at 249 GHz, revealing spatially resolved ring-like distribution in all tracers.
However, while continuum and CO isotopologues show asymmetric emission, with much stronger signal in the West, at radii 150 -- 180 au, SO is distributed further away from AB Aur, peaking near 250 au and is much more uniform in azimuth, only peaking slightly in the North-East and not on the dust trap, as already quoted by \cite{Pacheco+2016} and \cite{Riviere+2020}. 

All three SO lines have similar intensities, although the 6-5 lines are detected with a more limited signal-to-noise ratio. To better illustrate the detection, we present in Fig.\ref{fig:kepler} the integrated spectra and radial distributions obtained after correction from the Keplerian velocity field, using the \texttt{KEPLER} command from \textsc{Imager}. 
A comparison between the various radial brightness profiles (continuum, C$^{17}$O and SO lines) is shown in Fig.\ref{fig:radial}, while Fig.\ref{fig:azimut} shows the azimuthal dependencies for the continuum and SO. Both figures show that the SO emission is not specifically associated with the dust trap, SO actually peaking almost at an opposite azimuth compared to the continuum. Our SO observations agree with those of \cite{Riviere+2020} considering the lower signal-to-noise ratio of the latter.

\paragraph{Modeling} We use the \textsc{Diskfit} radiative transfer program \citep{Pietu+2007} to analyze the data in the Fourier plane. The continuum opacity is low. Hence, we use continuum-subtracted data for this purpose. All three SO lines were fitted simultaneously, allowing us to derive the rotation temperature of SO.  The model assumes LTE conditions and azimuthal symmetry. We assumed that the molecular column density and the rotation temperature follow power laws as a function of radius, with the molecule distribution truncated by inner and outer radii.  The vertical distribution of the molecules is assumed to follow a Gaussian distribution, with a scale height being a power law of radius. The local velocity dispersion (which includes thermal and turbulent motions on scales smaller than the beam size) is taken to be radially constant.
\begin{table}[!th] 
\tiny 
\caption{AB Aurigae ring parameters derived from SO lines}             
\label{tab:disk}      
\centering    
\begin{tabular}{|c c|l|}        
\hline              
\multicolumn{2}{|c|}{Geometry} & \multicolumn{1}{c|}{Law} \\ 
\hline 
Distance & 155.9 pc & \\
$V_\mathrm{LSR}$ & 5.85 km s$^{-1}$ & $V(r)=4.55\pm0.4\,(\frac{r}{100\,\textnormal{au}})^{-0.50\pm0.01}$ \, km\,s$^{-1}$ \\ 
&&\\
Inclination & 22$^\circ$ & $T(r)= 20 \pm 1\, (\frac{r}{200\,\textnormal{au}})^{-0.4}$ K \\ 
Orientation & -36.5$^\circ$  & \\ 
&&$H(r)= 11\, (\frac{r}{100\,\textnormal{au}})^{1.25}$ au \\
R$_{int}$ & 165$\pm$3 au &\\ 
R$_{out}$ & 400$\pm$10 au & $\Sigma(r)=1.5 \pm 0.1 \, 10^{13}\,(\frac{r}{200\,\textnormal{au}})^{-1.65\pm0.05}$ cm$^{-2}$ \\
& & \\
& & $\delta$V = $0.25 \pm 0.01$ \kms \\
\hline 
\end{tabular}
\vspace{-2.5ex}
\tablefoot{\small{Parameters with no error bars were fixed
at their specified values. $\delta$\,V is the FWHM local line width. The temperature exponent being loosely constrained ($0.54\pm0.20$) was assumed to be 0.4.}}
\label{tab:SO-model}
\end{table}
Table \ref{tab:SO-model} presents the best-fit model from the three SO lines.  Fig\,\ref{fig:so-composite} shows this model superimposed to the integrated spectrum of the SO $5_6-4_5$ line. The velocity index is found Keplerian ($0.50\pm0.01$). The best-fit models for the C$^{17}$O 2-1 transition are presented in Appendix B. 

Apart from a small shift in apparent PA of the rotation axis
(by $2-3^\circ$), the C$^{17}$O and SO determinations of the disk ``geometrical'' parameters (centre of rotation, orientation, inclination, systemic and rotation velocities) are consistent among them. The inclination derived from SO and C$^{17}$O, 22$^o$, is lower than the values reported by most previous mm observations, which were obtained with significantly lower angular resolutions and with more optically thick molecular tracers exhibiting
line-of-sight confusion with the surrounding cloud. With the new value
derived from these new unconfused observations, the large-scale outer ring appears almost coplanar with the inner disk surrounding the star \citep{Pantin+2005,Lazareff+2017}. The derived star mass is $2.4 \Msun$. 

Our fit also confirms that the SO emission is essentially located beyond the continuum ring \citep[as already quoted by][]{Riviere+2020} and extends up to 400 au contrary to the C$^{17}$O 2-1 emission, which mostly resides inside the SO ring, with inner and outer radii of $\sim$ 125 and 286 au, respectively (see Appendix \ref{app:co}). For comparison, the measured inner radius of CO is about $\sim 90$ au \citep{Pietu+2007} and that of the dust emission at 1.3\,mm is $\sim 140$ au \citep{Tang+2017}.

\section{Origin of SO emission}

\subsection{SO emission in the ring and the dust trap}


Sulfur-bearing species detected in the AB Aur ring are CS, SO \citep{Fuente+2010,Pacheco+2016} and H$_2$S \citep{Riviere+2022}. Contrary to H$_2$S, the SO emission does not peak at the location of the dense dust trap. The SO behaviour differs from what is observed around the warm source of IRS 48 where \cite{Booth+2021} reported SO emission is peaking at the dust trap. In this latter source, SO is expected to be a product of ice sublimation at the edge of the dust cavity, with the bulk of the ice reservoir located in the dust trap. On the other hand, SO rings are already detected around some Herbig Ae stars such as HD\,161942 \citep{Law+2023}, although the latter ring appears less symmetric than in the case of AB Aur. SO emission has also been reported in the Class I objects DG Tau and HL Tau \citep{Garufi+2022}, and in the evolved Class I source Oph16 IRS63 \citep{Flores+2023}. 

\paragraph{Gas temperature:} 
From detected SO lines, our best \textsc{Diskfit} model gives an average SO gas temperature of 20 K at 200 au. Since SO is not seen at the dust trap location, its temperature should be more characteristic of the ring temperature. On the contrary, C$^{17}$O emission is seen in the ring but peaks at the dust trap location (Fig.\ref{fig:areas}). The average temperature derived using DiskFit from  C$^{17}$O data is higher, 28 K, but this assumes azimuthal symmetry.  A more accurate derivation that takes advantage of the hyperfine structure of C$^{17}$O 2-1 (see Appendix \ref{app:co}) gives an average kinetic temperature above 20 K inside the ring, in good agreement with the temperature derived from SO.
The peak temperature of C$^{17}$O in the dust trap is within the range 35-42\,K (see Fig.\ref{fig:tk-C17O}), a somewhat high temperature which may reflect the fact that the C$^{17}$O emission does not come from the disk mid-plane but above due to its opacity. This deserves a more detailed modelling that is out of the scope of this letter. 
Nevertheless, these new temperature estimates are lower than those from \cite{Riviere+2020}, who derived an average temperature of 37-39 K inside the ring based on noisier observations.
 
\paragraph{Column density \& abundances:}
For SO, we derive an average surface density of $\sim 1.5 \cdot 10^{13}$ cm$^{-2}$ at 200 au  using \textsc{Diskfit}, in correct agreement with the $2.5 \cdot 10^{13}$ cm$^{-2}$ value quoted by \citet{Riviere+2020} from previous NOEMA observations. The detection of C$^{17}$O at temperatures of 20\,K or above
allows for a direct estimate of the H$_2$ column density since
at these temperatures, CO is not expected to be depleted. 
Using isotopic ratios $[^{16}$O$/^{18}$O$] = 557$ and $[^{18}$O$/^{17}$O$] = 3.6$ \citep{Wilson_1999}, the averaged surface
density derived using DiskFit modeling ($\sim 4\,10^{15}$\,cm$^{-2}$, see Appendix \ref{app:co}), and a normal [CO/H$_2$]
abundance ratio of $10^{-4}$, this yields an H$_2$ surface density of $8\,10^{22}$\,cm$^{-2}$, and thus a SO abundance
of  $\sim 1.7\,10^{-10}$.

Using the astro-chemical model \textsc{Nautilus}, \cite{Riviere+2022} performed a modelling of sulfur-bearing species in the AB Aurigae disk assuming no Sulfur depletion. They predicted a column density of $\sim 8 \cdot 10^{14}$ cm$^{-2}$, a factor $\sim 60$ higher than what we observe. Our results confirm that  SO is abundant in this disk as already mentioned by \cite{Riviere+2020}. SO emission was also observed in the disk of HD\,100546 \citep{Booth+2023}. As reported by \cite{Semenov+2018}, the observed low surface density of SO in disks, while CO is still abundant, suggests a relatively high C/O ratio in the disk molecular layers. Thus, our new observations deserve deeper chemical modelling to quantify this. 

\paragraph{Outer accretion streamers and shocks:}
Figure \ref{fig:radial} clearly shows that SO peaks at larger distances than C$^{17}$O and the dust emission. This fact is confirmed by the SO modelling (Table.\ref{tab:SO-model}). The inner radius of the SO ring is $\sim$160 au, and its outer radius is  400 au. An important fraction of the SO emission lies outside the dust and C$^{17}$O ring, questioning its origin. 
\cite{Tang+2012} reported the presence of several accretion streamers observed in CO 2-1 with the PdBI (now NOEMA), suggesting that the disk is still accreting material from outside. With the advent of ALMA, such streamers are now commonly observed around Class I objects (e.g. \cite{Sakai+2014}). It is tempting to explain the SO emission as resulting from slow shocks at the interface between the ring/disk and the outer streamers, SO being created locally and thus co-moving with the ring. A similar hypothesis is invoked to explain the observations of SO spirals in the outer disk of Oph16 IRS 63 \citep{Flores+2023} and the origin of the SO emission observed in DG Tau and HL Tau \citep{Garufi+2022}. \cite{VanGelder+2021} recently developed a model of slow shocks to explain molecular abundances specific to shocks observed at the disk-envelope interface of low-mass proto-stars. The physical conditions are similar to those found in the gas ring of AB Aur (a gas density in the range of $10^5-10^8$ cm$^{-3}$ and a low dark cloud temperature). Under standard assumptions (cosmic ray ionization rate of $10^{-17}$\,s$^{-1}$ and G$_o$ = 1) in low velocity ($\sim$ 2-3 km/s) shocks and cold medium ($\sim$ 20 K), the route to form SO starts with the desorption of CH$_4$ from grains, followed by the gas phase formation of CH$_3$ and H$_2$CO leading then to the formation of SH which reacts with O to form SO+H \citep [][their Fig.3 and 5]{VanGelder+2021}. Such a scenario, originally developed to explain accretion at the interface envelope-disk in Class I proto-stars, seems plausible in explaining the origin of the extended SO emission in the ring around AB Aur. Since the amount of shocked gas is partly limited by the accretion rate, the overall SO abundance relative to the whole disk mass may remain moderate. Detailed modelling is required to determine the shock amplitude and the expected amount of SO. 

\subsection{SO emission inside the cavity}

Figure \ref{fig:so-composite} presents the integrated spectrum of the main SO line (transition 5$_6$-4$_5$) on which the best fit derived using \textsc{Diskfit} is superimposed, and the 
integrated spectrum is dominated by the emission of the ring (Fig.\ref{fig:so-composite}), middle bottom panel. Interestingly, there is a red-shifted wing (Fig.\ref{fig:so-composite}, right bottom panel) in the velocity range of 7.4-8.5 km/s with no symmetric counterpart in the blue-shifted side (Fig.\ref{fig:so-composite}, left bottom panel). The images of the wings have been obtained by imaging and CLEANing the visibilities only in their specific velocity ranges. The red wing emission peaks at about 5 sigma and is located close to the inner disk of AB Aur and the candidate proto-planet P1 (shown as a red point).

A first possibility would be that this SO emission mostly originates from the inner disk of AB Aurigae. In this case, a symmetric blue-shifted wing would be expected, but the lack of sensitivity might preclude its detection. 
We also note that the emission region
is close to the inner ends of the S3 and S4 outer spirals identified by \citet{Tang+2012}.

Another possibility would be that the redshifted SO emission is linked to the presence of the embedded proto-planet P1. Based on \cite{Boccaletti+2020}, the Keplerian velocity of P1 is about 2.5 km/s. Adding the systemic velocity (5.8 km/s), we get a projected velocity of 8.3 km/s, which is in agreement with the velocity range that is observed, even taking into account the uncertainties, particularly on the inclination. This suggests that, at least, part of this SO emission may originate close to the hot spot observed both in CO and in NIR. In this hypothesis, the SO emission could trace a streamer-like feature that would fall onto the (undetected) circumplanetary disk of P1. However, the limited angular resolution and signal to noise prevents to reveal the velocity gradient what would be expected along the streamer.

Note that such possible links between embedded planets and SO emission have already been seen in similar systems. From ALMA observations, \cite{Law+2023} reported SO and SiS emissions at the location of a putative planet of 2$M_{Jup}$ orbiting in the cavity of HD\,169142 while \cite{Booth+2023} reported SO emission inside the cavity around HD\,100546, potentially connected to an embedded planet. Hence, the observation of SO in AB Aur at the location of the embedded planet P1 would be the third case of indirect detection of a circumplanetary disk. 

\section{Summary}

Using ALMA archival data of the AB Aurigae system, we image the SO emission in the close
environment of the star. Our analysis reveals that: 
\begin{itemize}
\item The SO emission is partly located inside the molecular (CO) ring but does not peak at the location of the dust trap where the gas temperature, estimated from C$^{17}$O hyperfine components, is within 35 - 40 K, a factor 2 warmer than in the SO ring (20 K). 
\item Surprisingly, the inner radius of the SO ring is at $\sim 160$ au, while the CO ring has an inner radius of about 90 au. In agreement with slow velocity accretion shock models in proto-stars, SO could be formed locally in the gas phase in the accretion shocks generated at the interface between the observed outer CO spirals and the molecular and dust ring. 
\item We also observed unresolved red-shifted SO emission at a position in agreement with the embedded planet candidate P1 (within $\sim 30$ au) from the star. The observed velocity of the SO emission is compatible with the expected Keplerian velocity at P1. A possibility would be that SO traces an accretion shock between the circumplanetary disk of P1 and its surroundings.
\item The SO emission being weak in these objects, long integration times with ALMA are needed to better characterize the link between SO, a tracer of accretion shock, and embedded planets in gas-rich young planetary systems. 
\end{itemize}

\begin{acknowledgements}
 This work was supported by ``Programme National de Physique Stellaire'' (PNPS) and ``Programme National de Physique Chimie du Milieu Interstellaire'' (PCMI) from INSU/CNRS.
This research made use of the SIMBAD database, operated at the CDS, Strasbourg, France.
This work was partially supported by the ORCHID program (project number: 49523PG), funded by the French Ministry for Europe and Foreign Affairs, the French Ministry for Higher Education and Research and the National Science and Technology Council (NSTC)
AD warmly thanks the ESO ALMA team which has performed the calibration of the archival MS. 
Y.W.T. acknowledges support through NSTC grant 111-2112-M-001-064- and 112-2112-M-001-066-.
A.F. and P.R.M. are members of project PID2022-137980NB-I00, funded by MCIN/AEI/10.13039/501100011033/FEDER UE.
This paper makes use of the following ALMA data: ADS/JAO.ALMA\#2019.1.00579.S, ADS/JAO.ALMA\#2021.1.00690.S and ADS/JAO.ALMA\#2021.1.01216.S. ALMA is a partnership of ESO (representing its member states), NSF (USA) and NINS (Japan), together with NRC (Canada), NSTC and ASIAA (Taiwan), and KASI (Republic of Korea), in cooperation with the Republic of Chile. The Joint ALMA Observatory is operated by ESO, AUI/NRAO and NAOJ.
\end{acknowledgements}


\begin{thebibliography}{45}
\expandafter\ifx\csname natexlab\endcsname\relax\def\natexlab#1{#1}\fi

\bibitem[{{Baruteau} {et~al.}(2021){Baruteau}, {Wafflard-Fernandez}, {Le Gal},
  {Debras}, {Carmona}, {Fuente}, \& {Rivi{\`e}re-Marichalar}}]{Baruteau+2021}
{Baruteau}, C., {Wafflard-Fernandez}, G., {Le Gal}, R., {et~al.} 2021, \mnras,
  505, 359

\bibitem[{{Benisty} {et~al.}(2015){Benisty}, {Juhasz}, {Boccaletti},
  {Avenhaus}, {Milli}, {Thalmann}, {Dominik}, {Pinilla}, {Buenzli}, {Pohl},
  {Beuzit}, {Birnstiel}, {de Boer}, {Bonnefoy}, {Chauvin}, {Christiaens},
  {Garufi}, {Grady}, {Henning}, {Huelamo}, {Isella}, {Langlois}, {M{\'e}nard},
  {Mouillet}, {Olofsson}, {Pantin}, {Pinte}, \& {Pueyo}}]{Benisty+2015}
{Benisty}, M., {Juhasz}, A., {Boccaletti}, A., {et~al.} 2015, \aap, 578, L6

\bibitem[{{Boccaletti} {et~al.}(2020){Boccaletti}, {Di Folco}, {Pantin},
  {Dutrey}, {Guilloteau}, {Tang}, {Pi{\'e}tu}, {Habart}, {Milli}, {Beck}, \&
  {Maire}}]{Boccaletti+2020}
{Boccaletti}, A., {Di Folco}, E., {Pantin}, E., {et~al.} 2020, \aap, 637, L5

\bibitem[{{Booth} {et~al.}(2023){Booth}, {Ilee}, {Walsh}, {Kama}, {Keyte}, {van
  Dishoeck}, \& {Nomura}}]{Booth+2023}
{Booth}, A.~S., {Ilee}, J.~D., {Walsh}, C., {et~al.} 2023, \aap, 669, A53

\bibitem[{{Booth} {et~al.}(2021){Booth}, {van der Marel}, {Leemker}, {van
  Dishoeck}, \& {Ohashi}}]{Booth+2021}
{Booth}, A.~S., {van der Marel}, N., {Leemker}, M., {van Dishoeck}, E.~F., \&
  {Ohashi}, S. 2021, \aap, 651, L6

\bibitem[{{Corder} {et~al.}(2005){Corder}, {Eisner}, \&
  {Sargent}}]{Corder+2005}
{Corder}, S., {Eisner}, J., \& {Sargent}, A. 2005, \apjl, 622, L133

\bibitem[{{Currie} {et~al.}(2022){Currie}, {Lawson}, {Schneider}, {Lyra},
  {Wisniewski}, {Grady}, {Guyon}, {Tamura}, {Kotani}, {Kawahara}, {Brandt},
  {Uyama}, {Muto}, {Dong}, {Kudo}, {Hashimoto}, {Fukagawa}, {Wagner}, {Lozi},
  {Chilcote}, {Tobin}, {Groff}, {Ward-Duong}, {Januszewski}, {Norris},
  {Tuthill}, {van der Marel}, {Sitko}, {Deo}, {Vievard}, {Jovanovic},
  {Martinache}, \& {Skaf}}]{Currie+2022}
{Currie}, T., {Lawson}, K., {Schneider}, G., {et~al.} 2022, Nature Astronomy,
  6, 751

\bibitem[{{Dong} {et~al.}(2015){Dong}, {Zhu}, {Rafikov}, \&
  {Stone}}]{Dong+2015}
{Dong}, R., {Zhu}, Z., {Rafikov}, R.~R., \& {Stone}, J.~M. 2015, \apjl, 809, L5

\bibitem[{{Flores} {et~al.}(2023){Flores}, {Ohashi}, {Tobin}, {J{\o}rgensen},
  {Takakuwa}, {Li}, {Lin}, {van't Hoff}, {Plunkett}, {Yamato}, {Sai (Insa
  Choi)}, {Koch}, {Yen}, {Aikawa}, {Aso}, {de Gregorio-Monsalvo}, {Kido},
  {Kwon}, {Lee}, {Lee}, {Looney}, {Santamar{\'\i}a-Miranda}, {Sharma},
  {Thieme}, {Williams}, {Han}, {Narayanan}, \& {Lai}}]{Flores+2023}
{Flores}, C., {Ohashi}, N., {Tobin}, J.~J., {et~al.} 2023, \apj, 958, 98

\bibitem[{{Fuente} {et~al.}(2017){Fuente}, {Baruteau}, {Neri}, {Carmona},
  {Ag{\'u}ndez}, {Goicoechea}, {Bachiller}, {Cernicharo}, \&
  {Bern{\'e}}}]{Fuente+2017}
{Fuente}, A., {Baruteau}, C., {Neri}, R., {et~al.} 2017, \apjl, 846, L3

\bibitem[{{Fuente} {et~al.}(2010){Fuente}, {Cernicharo}, {Ag{\'u}ndez},
  {Bern{\'e}}, {Goicoechea}, {Alonso-Albi}, \& {Marcelino}}]{Fuente+2010}
{Fuente}, A., {Cernicharo}, J., {Ag{\'u}ndez}, M., {et~al.} 2010, \aap, 524,
  A19

\bibitem[{{Fukagawa} {et~al.}(2004){Fukagawa}, {Hayashi}, {Tamura}, {Itoh},
  {Hayashi}, {Oasa}, {Takeuchi}, {Morino}, {Murakawa}, {Oya}, {Yamashita},
  {Suto}, {Mayama}, {Naoi}, {Ishii}, {Pyo}, {Nishikawa}, {Takato}, {Usuda},
  {Ando}, {Iye}, {Miyama}, \& {Kaifu}}]{Fukagawa+2004}
{Fukagawa}, M., {Hayashi}, M., {Tamura}, M., {et~al.} 2004, \apjl, 605, L53

\bibitem[{{Fukagawa} {et~al.}(2006){Fukagawa}, {Tamura}, {Itoh}, {Kudo},
  {Imaeda}, {Oasa}, {Hayashi}, \& {Hayashi}}]{Fukagawa+2006}
{Fukagawa}, M., {Tamura}, M., {Itoh}, Y., {et~al.} 2006, \apjl, 636, L153

\bibitem[{{Gaia Collaboration}(2022)}]{Gaia+2022}
{Gaia Collaboration}. 2022, {VizieR Online Data Catalog: Gaia DR3 Part 1. Main
  source (Gaia Collaboration, 2022)}, VizieR On-line Data Catalog: I/355.
  Originally published in: Astron. Astrophys., in prep. (2022)

\bibitem[{{Garufi} {et~al.}(2022){Garufi}, {Podio}, {Codella}, {Segura-Cox},
  {Vander Donckt}, {Mercimek}, {Bacciotti}, {Fedele}, {Kasper}, {Pineda},
  {Humphreys}, \& {Testi}}]{Garufi+2022}
{Garufi}, A., {Podio}, L., {Codella}, C., {et~al.} 2022, \aap, 658, A104

\bibitem[{{Garufi} {et~al.}(2013){Garufi}, {Quanz}, {Avenhaus}, {Buenzli},
  {Dominik}, {Meru}, {Meyer}, {Pinilla}, {Schmid}, \& {Wolf}}]{Garufi+2013}
{Garufi}, A., {Quanz}, S.~P., {Avenhaus}, H., {et~al.} 2013, \aap, 560, A105

\bibitem[{{Grady} {et~al.}(2013){Grady}, {Muto}, {Hashimoto}, {Fukagawa},
  {Currie}, {Biller}, {Thalmann}, {Sitko}, {Russell}, {Wisniewski}, {Dong},
  {Kwon}, {Sai}, {Hornbeck}, {Schneider}, {Hines}, {Moro Mart{\'\i}n}, {Feldt},
  {Henning}, {Pott}, {Bonnefoy}, {Bouwman}, {Lacour}, {Mueller}, {Juh{\'a}sz},
  {Crida}, {Chauvin}, {Andrews}, {Wilner}, {Kraus}, {Dahm}, {Robitaille},
  {Jang-Condell}, {Abe}, {Akiyama}, {Brandner}, {Brandt}, {Carson}, {Egner},
  {Follette}, {Goto}, {Guyon}, {Hayano}, {Hayashi}, {Hayashi}, {Hodapp},
  {Ishii}, {Iye}, {Janson}, {Kandori}, {Knapp}, {Kudo}, {Kusakabe}, {Kuzuhara},
  {Mayama}, {McElwain}, {Matsuo}, {Miyama}, {Morino}, {Nishimura}, {Pyo},
  {Serabyn}, {Suto}, {Suzuki}, {Takami}, {Takato}, {Terada}, {Tomono},
  {Turner}, {Watanabe}, {Yamada}, {Takami}, {Usuda}, \& {Tamura}}]{Grady+2013}
{Grady}, C.~A., {Muto}, T., {Hashimoto}, J., {et~al.} 2013, \apj, 762, 48

\bibitem[{{Grady} {et~al.}(1999){Grady}, {Woodgate}, {Bruhweiler}, {Boggess},
  {Plait}, {Lindler}, {Clampin}, \& {Kalas}}]{Grady+1999}
{Grady}, C.~A., {Woodgate}, B., {Bruhweiler}, F.~C., {et~al.} 1999, \apjl, 523,
  L151

\bibitem[{{Haffert} {et~al.}(2019){Haffert}, {Bohn}, {de Boer}, {Snellen},
  {Brinchmann}, {Girard}, {Keller}, \& {Bacon}}]{Haffert+2019}
{Haffert}, S.~Y., {Bohn}, A.~J., {de Boer}, J., {et~al.} 2019, Nature
  Astronomy, 3, 749

\bibitem[{{Isella} {et~al.}(2019){Isella}, {Benisty}, {Teague}, {Bae},
  {Keppler}, {Facchini}, \& {P{\'e}rez}}]{Isella+2019}
{Isella}, A., {Benisty}, M., {Teague}, R., {et~al.} 2019, \apjl, 879, L25

\bibitem[{{Isella} {et~al.}(2010){Isella}, {Natta}, {Wilner}, {Carpenter}, \&
  {Testi}}]{Isella+2010}
{Isella}, A., {Natta}, A., {Wilner}, D., {Carpenter}, J.~M., \& {Testi}, L.
  2010, \apj, 725, 1735

\bibitem[{{Keppler} {et~al.}(2018){Keppler}, {Benisty}, {M{\"u}ller},
  {Henning}, {van Boekel}, {Cantalloube}, {Ginski}, {van Holstein}, {Maire},
  {Pohl}, {Samland}, {Avenhaus}, {Baudino}, {Boccaletti}, {de Boer},
  {Bonnefoy}, {Chauvin}, {Desidera}, {Langlois}, {Lazzoni}, {Marleau},
  {Mordasini}, {Pawellek}, {Stolker}, {Vigan}, {Zurlo}, {Birnstiel},
  {Brandner}, {Feldt}, {Flock}, {Girard}, {Gratton}, {Hagelberg}, {Isella},
  {Janson}, {Juhasz}, {Kemmer}, {Kral}, {Lagrange}, {Launhardt}, {Matter},
  {M{\'e}nard}, {Milli}, {Molli{\`e}re}, {Olofsson}, {P{\'e}rez}, {Pinilla},
  {Pinte}, {Quanz}, {Schmidt}, {Udry}, {Wahhaj}, {Williams}, {Buenzli},
  {Cudel}, {Dominik}, {Galicher}, {Kasper}, {Lannier}, {Mesa}, {Mouillet},
  {Peretti}, {Perrot}, {Salter}, {Sissa}, {Wildi}, {Abe}, {Antichi},
  {Augereau}, {Baruffolo}, {Baudoz}, {Bazzon}, {Beuzit}, {Blanchard}, {Brems},
  {Buey}, {De Caprio}, {Carbillet}, {Carle}, {Cascone}, {Cheetham}, {Claudi},
  {Costille}, {Delboulb{\'e}}, {Dohlen}, {Fantinel}, {Feautrier}, {Fusco},
  {Giro}, {Gluck}, {Gry}, {Hubin}, {Hugot}, {Jaquet}, {Le Mignant}, {Llored},
  {Madec}, {Magnard}, {Martinez}, {Maurel}, {Meyer}, {M{\"o}ller-Nilsson},
  {Moulin}, {Mugnier}, {Orign{\'e}}, {Pavlov}, {Perret}, {Petit}, {Pragt},
  {Puget}, {Rabou}, {Ramos}, {Rigal}, {Rochat}, {Roelfsema}, {Rousset}, {Roux},
  {Salasnich}, {Sauvage}, {Sevin}, {Soenke}, {Stadler}, {Suarez}, {Turatto}, \&
  {Weber}}]{Keppler+2018}
{Keppler}, M., {Benisty}, M., {M{\"u}ller}, A., {et~al.} 2018, \aap, 617, A44

\bibitem[{{Law} {et~al.}(2023){Law}, {Booth}, \& {{\"O}berg}}]{Law+2023}
{Law}, C.~J., {Booth}, A.~S., \& {{\"O}berg}, K.~I. 2023, \apjl, 952, L19

\bibitem[{{Lazareff} {et~al.}(2017){Lazareff}, {Berger}, {Kluska}, {Le
  Bouquin}, {Benisty}, {Malbet}, {Koen}, {Pinte}, {Thi}, {Absil}, {Baron},
  {Delboulb{\'e}}, {Duvert}, {Isella}, {Jocou}, {Juhasz}, {Kraus}, {Lachaume},
  {M{\'e}nard}, {Millan-Gabet}, {Monnier}, {Moulin}, {Perraut}, {Rochat},
  {Soulez}, {Tallon}, {Thi{\'e}baut}, {Traub}, \& {Zins}}]{Lazareff+2017}
{Lazareff}, B., {Berger}, J.~P., {Kluska}, J., {et~al.} 2017, \aap, 599, A85

\bibitem[{{Muto} {et~al.}(2012){Muto}, {Grady}, {Hashimoto}, {Fukagawa},
  {Hornbeck}, {Sitko}, {Russell}, {Werren}, {Cur{\'e}}, {Currie}, {Ohashi},
  {Okamoto}, {Momose}, {Honda}, {Inutsuka}, {Takeuchi}, {Dong}, {Abe},
  {Brandner}, {Brandt}, {Carson}, {Egner}, {Feldt}, {Fukue}, {Goto}, {Guyon},
  {Hayano}, {Hayashi}, {Hayashi}, {Henning}, {Hodapp}, {Ishii}, {Iye},
  {Janson}, {Kandori}, {Knapp}, {Kudo}, {Kusakabe}, {Kuzuhara}, {Matsuo},
  {Mayama}, {McElwain}, {Miyama}, {Morino}, {Moro-Martin}, {Nishimura}, {Pyo},
  {Serabyn}, {Suto}, {Suzuki}, {Takami}, {Takato}, {Terada}, {Thalmann},
  {Tomono}, {Turner}, {Watanabe}, {Wisniewski}, {Yamada}, {Takami}, {Usuda}, \&
  {Tamura}}]{Muto+2012ApJ}
{Muto}, T., {Grady}, C.~A., {Hashimoto}, J., {et~al.} 2012, \apjl, 748, L22

\bibitem[{{Pacheco-V{\'a}zquez} {et~al.}(2016){Pacheco-V{\'a}zquez}, {Fuente},
  {Baruteau}, {Bern{\'e}}, {Ag{\'u}ndez}, {Neri}, {Goicoechea}, {Cernicharo},
  \& {Bachiller}}]{Pacheco+2016}
{Pacheco-V{\'a}zquez}, S., {Fuente}, A., {Baruteau}, C., {et~al.} 2016, \aap,
  589, A60

\bibitem[{{Pantin} {et~al.}(2005){Pantin}, {Bouwman}, \&
  {Lagage}}]{Pantin+2005}
{Pantin}, E., {Bouwman}, J., \& {Lagage}, P.~O. 2005, \aap, 437, 525

\bibitem[{{P{\'e}rez} {et~al.}(2016){P{\'e}rez}, {Carpenter}, {Andrews},
  {Ricci}, {Isella}, {Linz}, {Sargent}, {Wilner}, {Henning}, {Deller},
  {Chandler}, {Dullemond}, {Lazio}, {Menten}, {Corder}, {Storm}, {Testi},
  {Tazzari}, {Kwon}, {Calvet}, {Greaves}, {Harris}, \& {Mundy}}]{Perez+2016}
{P{\'e}rez}, L.~M., {Carpenter}, J.~M., {Andrews}, S.~M., {et~al.} 2016,
  Science, 353, 1519

\bibitem[{{Perez} {et~al.}(2015){Perez}, {Dunhill}, {Casassus}, {Roman},
  {Szul{\'a}gyi}, {Flores}, {Marino}, \& {Montesinos}}]{Perez+2015}
{Perez}, S., {Dunhill}, A., {Casassus}, S., {et~al.} 2015, \apjl, 811, L5

\bibitem[{{Pi{\'e}tu} {et~al.}(2007){Pi{\'e}tu}, {Dutrey}, \&
  {Guilloteau}}]{Pietu+2007}
{Pi{\'e}tu}, V., {Dutrey}, A., \& {Guilloteau}, S. 2007, \aap, 467, 163

\bibitem[{{Pi{\'e}tu} {et~al.}(2005){Pi{\'e}tu}, {Guilloteau}, \&
  {Dutrey}}]{Pietu+2005}
{Pi{\'e}tu}, V., {Guilloteau}, S., \& {Dutrey}, A. 2005, \aap, 443, 945

\bibitem[{{Pinte} {et~al.}(2018){Pinte}, {Price}, {M{\'e}nard}, {Duch{\^e}ne},
  {Dent}, {Hill}, {de Gregorio-Monsalvo}, {Hales}, \& {Mentiplay}}]{Pinte+2018}
{Pinte}, C., {Price}, D.~J., {M{\'e}nard}, F., {et~al.} 2018, \apjl, 860, L13

\bibitem[{{Prasad} \& {Huntress}(1980)}]{Prasad+1980}
{Prasad}, S.~S. \& {Huntress}, W.~T., J. 1980, \apj, 239, 151

\bibitem[{{Rivi{\`e}re-Marichalar} {et~al.}(2022){Rivi{\`e}re-Marichalar},
  {Fuente}, {Esplugues}, {Wakelam}, {le Gal}, {Baruteau}, {Ribas},
  {Mac{\'\i}as}, {Neri}, \& {Navarro-Almaida}}]{Riviere+2022}
{Rivi{\`e}re-Marichalar}, P., {Fuente}, A., {Esplugues}, G., {et~al.} 2022,
  \aap, 665, A61

\bibitem[{{Rivi{\`e}re-Marichalar} {et~al.}(2020){Rivi{\`e}re-Marichalar},
  {Fuente}, {Le Gal}, {Baruteau}, {Neri}, {Navarro-Almaida},
  {Trevi{\~n}o-Morales}, {Mac{\'\i}as}, {Bachiller}, \&
  {Osorio}}]{Riviere+2020}
{Rivi{\`e}re-Marichalar}, P., {Fuente}, A., {Le Gal}, R., {et~al.} 2020, \aap,
  642, A32

\bibitem[{{Sakai} {et~al.}(2014){Sakai}, {Sakai}, {Hirota}, {Watanabe},
  {Ceccarelli}, {Kahane}, {Bottinelli}, {Caux}, {Demyk}, {Vastel}, {Coutens},
  {Taquet}, {Ohashi}, {Takakuwa}, {Yen}, {Aikawa}, \& {Yamamoto}}]{Sakai+2014}
{Sakai}, N., {Sakai}, T., {Hirota}, T., {et~al.} 2014, \nat, 507, 78

\bibitem[{{Semenov} {et~al.}(2018){Semenov}, {Favre}, {Fedele}, {Guilloteau},
  {Teague}, {Henning}, {Dutrey}, {Chapillon}, {Hersant}, \&
  {Pi{\'e}tu}}]{Semenov+2018}
{Semenov}, D., {Favre}, C., {Fedele}, D., {et~al.} 2018, \aap, 617, A28

\bibitem[{{Tang} {et~al.}(2017){Tang}, {Guilloteau}, {Dutrey}, {Muto}, {Shen},
  {Gu}, {Inutsuka}, {Momose}, {Pietu}, {Fukagawa}, {Chapillon}, {Ho}, {di
  Folco}, {Corder}, {Ohashi}, \& {Hashimoto}}]{Tang+2017}
{Tang}, Y.-W., {Guilloteau}, S., {Dutrey}, A., {et~al.} 2017, \apj, 840, 32

\bibitem[{{Tang} {et~al.}(2012){Tang}, {Guilloteau}, {Pi{\'e}tu}, {Dutrey},
  {Ohashi}, \& {Ho}}]{Tang+2012}
{Tang}, Y.~W., {Guilloteau}, S., {Pi{\'e}tu}, V., {et~al.} 2012, \aap, 547, A84

\bibitem[{{van Gelder} {et~al.}(2021){van Gelder}, {Tabone}, {van Dishoeck}, \&
  {Godard}}]{VanGelder+2021}
{van Gelder}, M.~L., {Tabone}, B., {van Dishoeck}, E.~F., \& {Godard}, B. 2021,
  \aap, 653, A159

\bibitem[{{Wilson}(1999)}]{Wilson_1999}
{Wilson}, T.~L. 1999, Reports on Progress in Physics, 62, 143

\bibitem[{{Zhou} {et~al.}(2021){Zhou}, {Bowler}, {Wagner}, {Schneider}, {Apai},
  {Kraus}, {Close}, {Herczeg}, \& {Fang}}]{Zhou+2021}
{Zhou}, Y., {Bowler}, B.~P., {Wagner}, K.~R., {et~al.} 2021, \aj, 161, 244

\bibitem[{{Zhou} {et~al.}(2023){Zhou}, {Bowler}, {Yang}, {Sanghi}, {Herczeg},
  {Kraus}, {Bae}, {Long}, {Follette}, {Ward-Duong}, {Zhu}, {Biddle}, {Close},
  {Jiang}, \& {Wu}}]{Zhou+2023}
{Zhou}, Y., {Bowler}, B.~P., {Yang}, H., {et~al.} 2023, \aj, 166, 220

\bibitem[{{Zhou} {et~al.}(2022){Zhou}, {Sanghi}, {Bowler}, {Wu}, {Close},
  {Long}, {Ward-Duong}, {Zhu}, {Kraus}, {Follette}, \& {Bae}}]{Zhou+2022}
{Zhou}, Y., {Sanghi}, A., {Bowler}, B.~P., {et~al.} 2022, \apjl, 934, L13

\bibitem[{{Zhu} {et~al.}(2015){Zhu}, {Dong}, {Stone}, \& {Rafikov}}]{Zhu+2015}
{Zhu}, Z., {Dong}, R., {Stone}, J.~M., \& {Rafikov}, R.~R. 2015, \apj, 813, 88

\end{thebibliography}
%
\begin{appendix}
\section{Complementary Figures}

\begin{figure*}
\centering
\includegraphics[width=14.3cm]{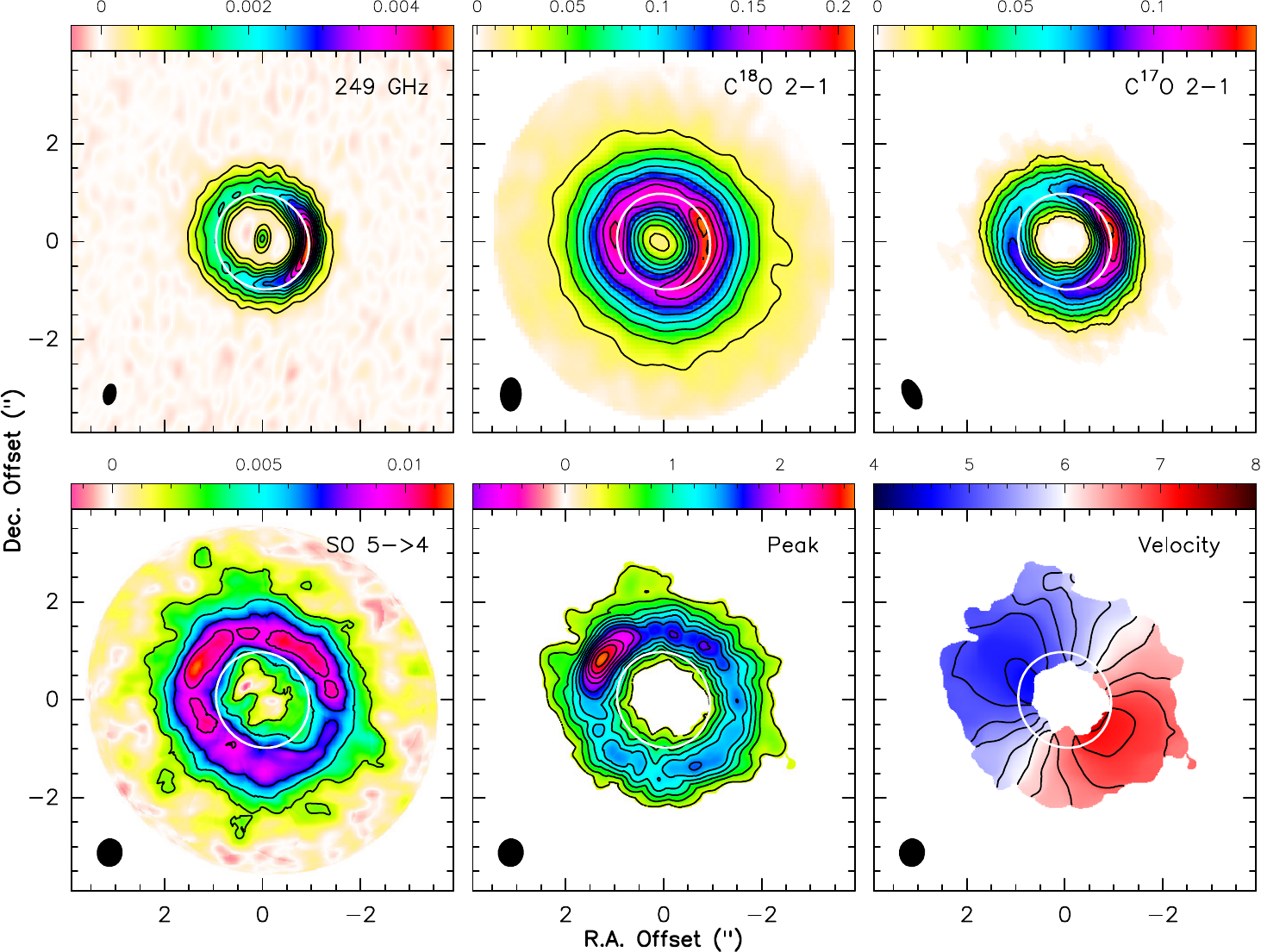}
\caption{Line and continuum observations. 
Top panels: from left to right - continuum image at 249 GHz (in mJy/beam), C$^{18}$O and C$^{17}$O J=2-1 integrated areas (in mJy/beam.km/s).
Bottom panels: SO $5_6-4_5$ (brighter) line, from left to right: integrated area, peak brightness (in K) and velocity field (in km/s).
Synthesized beams 
are indicated.
The white ellipse marks the approximate peak radius of the dust ring.
\label{fig:areas}}
\end{figure*}

\begin{figure*}
 \centering
\includegraphics[width=14.3cm]{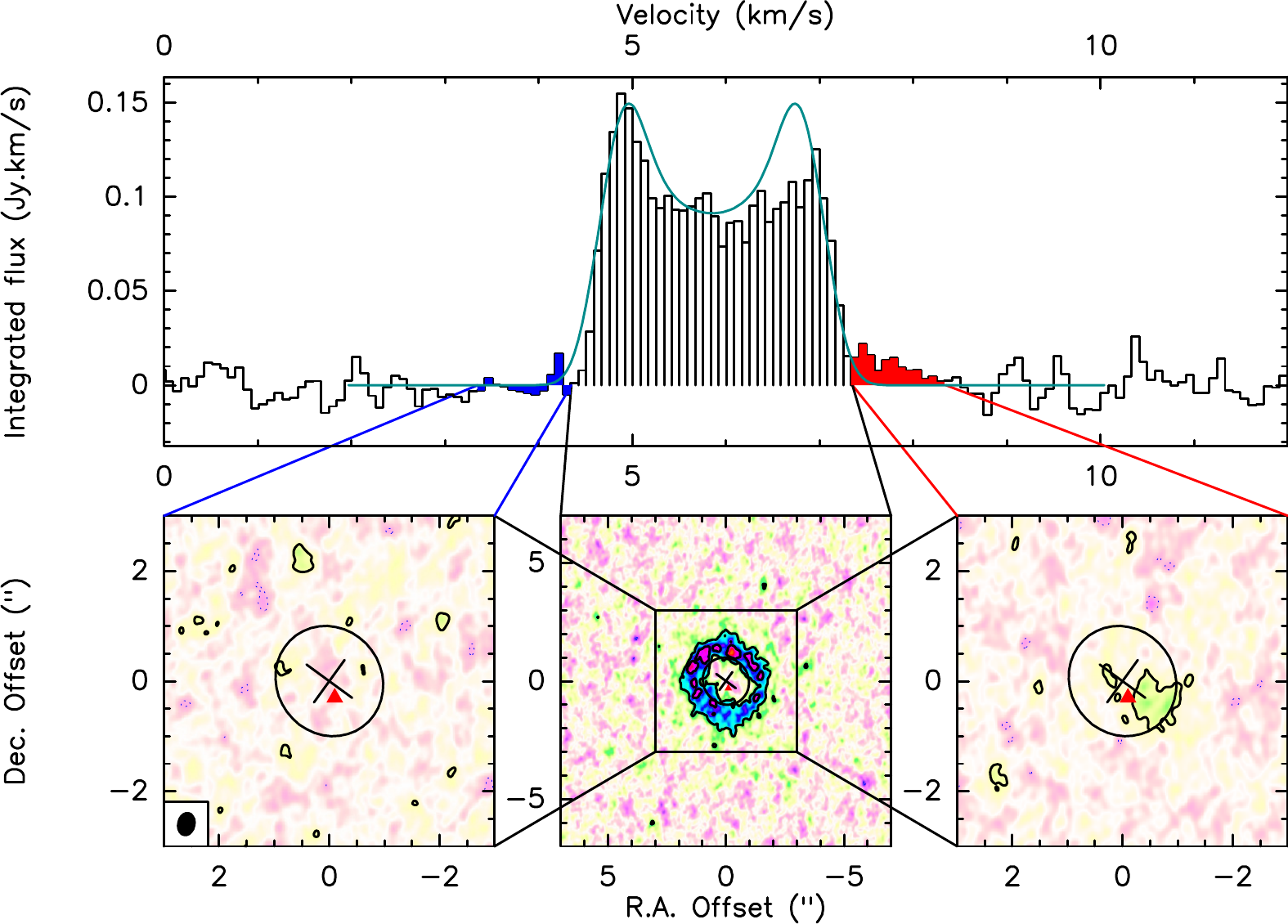}
\caption{SO $5_6-4_5$ towards AB Aur. Top: integrated spectrum (histogram), with best fit rotationally symmetric Keplerian
disk model superimposed (greenish curve). Bottom, from left to right: integrated emission over the indicated velocity ranges, from blueshifted
to redshifted. Contour levels are 2.5 $\sigma$  and 5 $\sigma$. The beam size is displayed in the left panel. The cross indicates
the AB Aur star position and the disk orientation (PA $126^\circ$) and inclination ($22^\circ$) and the ellipse marks the approximate peak radius of the dust
ring. The red triangle is the
approximate position of the putative proto-planet traced by CO and near-IR emission. }
\label{fig:so-composite}%
\end{figure*}

Figure \ref{fig:areas} presents the line and continuum images. Figure \ref{fig:so-composite} is a montage showing the SO emission and its best model. 

\section{Keplerian corrected profiles}
\label{app:kepler}
\begin{figure}
\centering
\includegraphics[width=0.9\columnwidth]{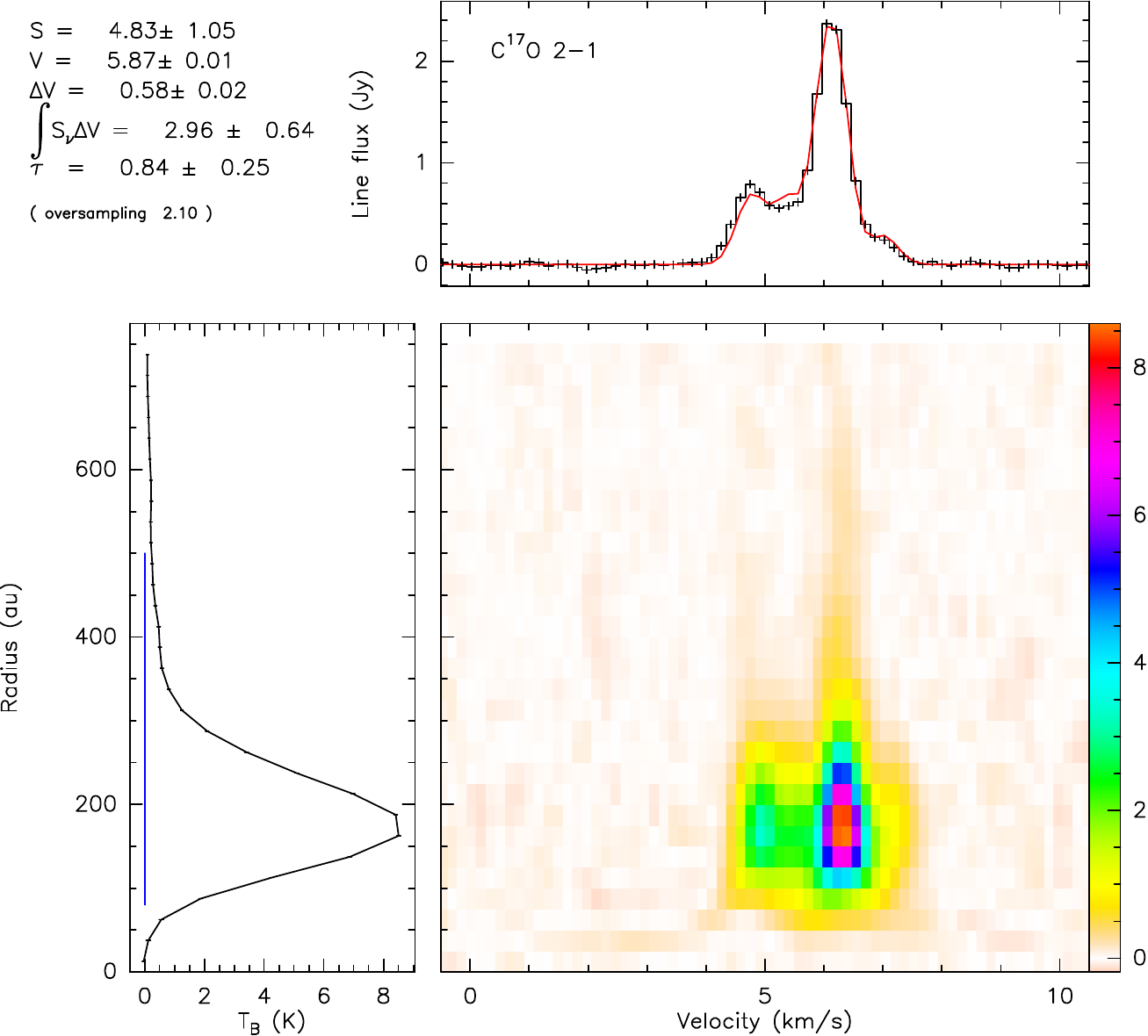}\\
\includegraphics[width=0.9\columnwidth]{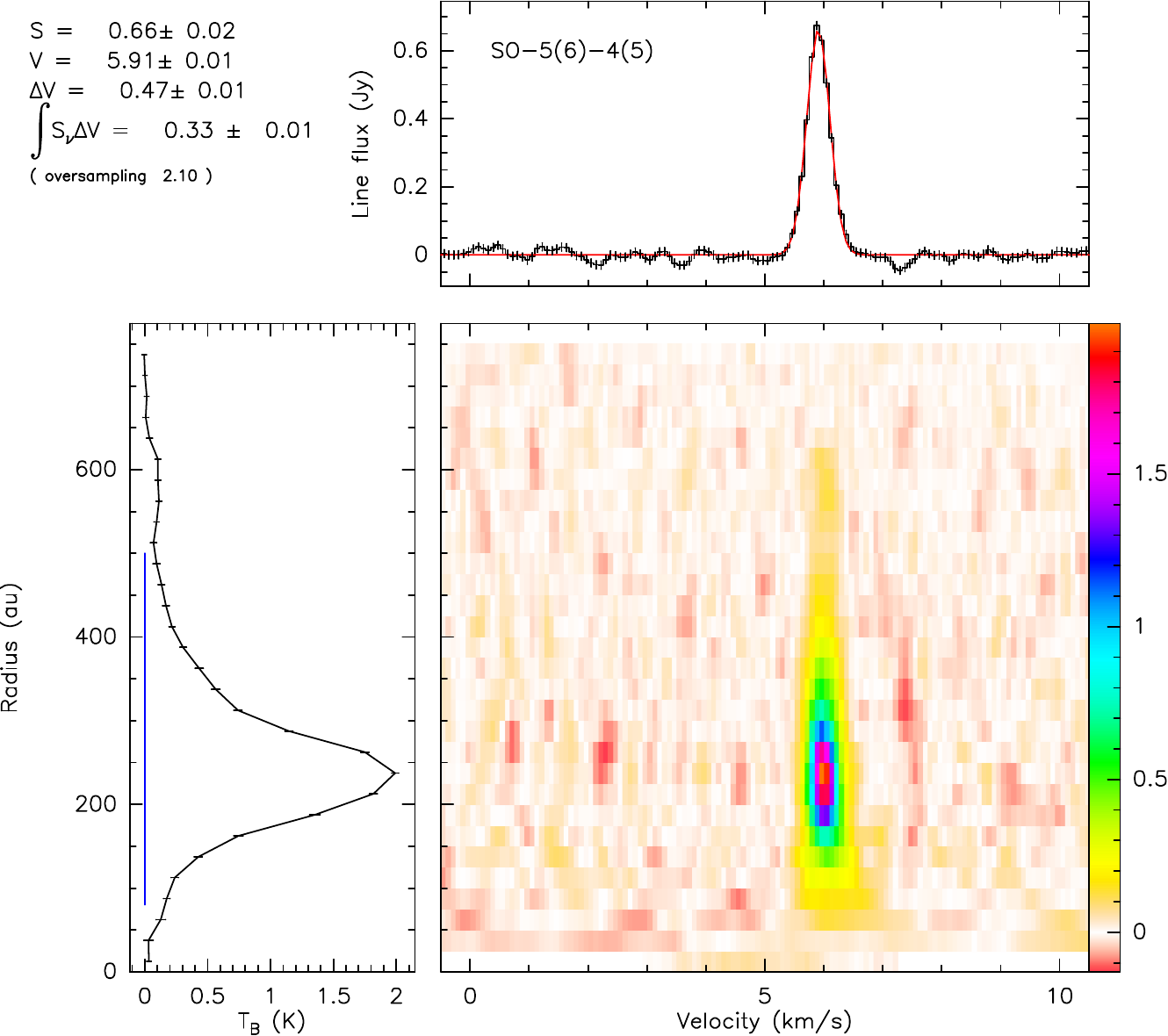}\\
\includegraphics[width=0.9\columnwidth]{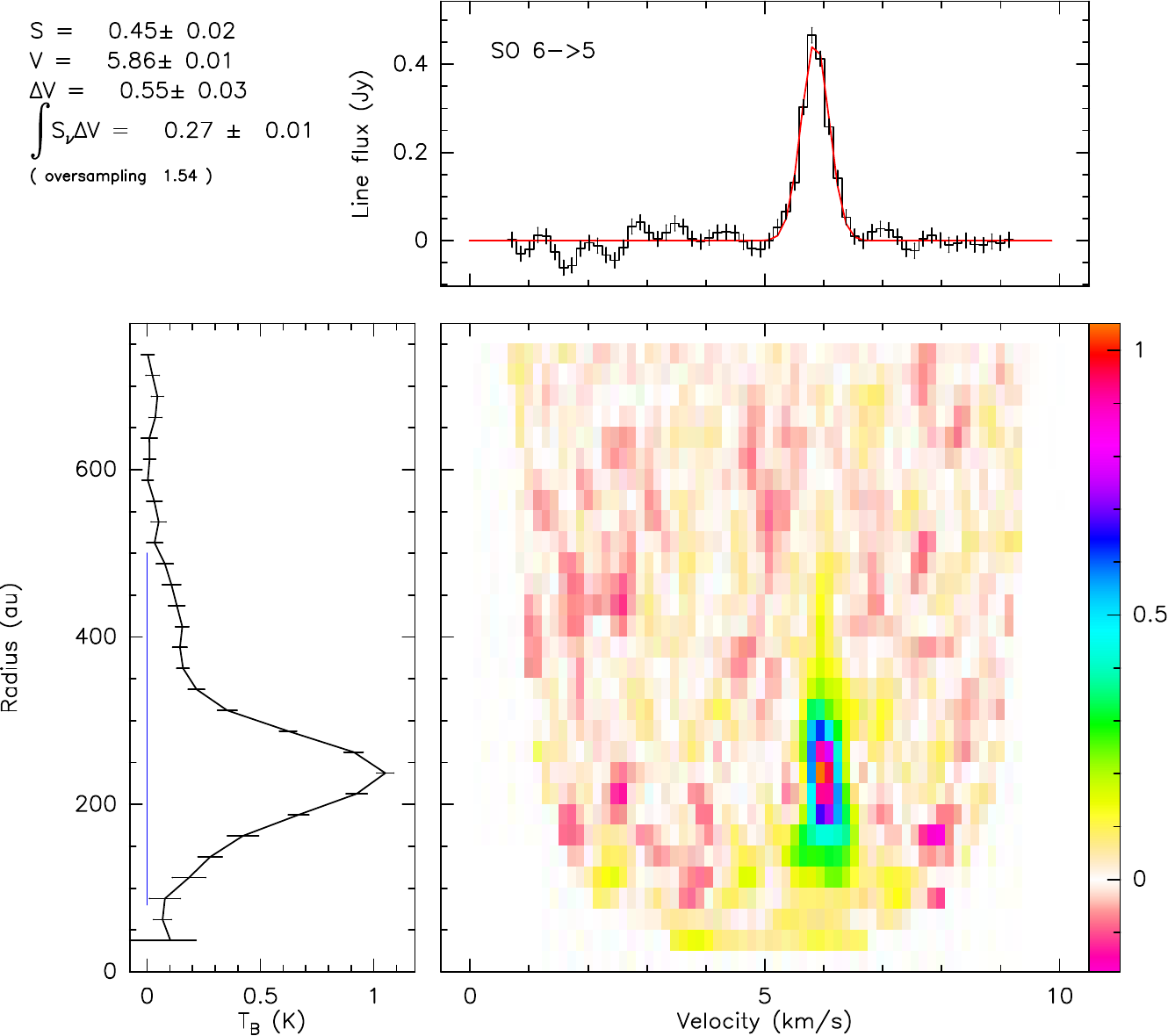}
\caption{Azimuthal average of line profiles obtained after correction from Keplerian rotation. 
The SO 6-5 data was obtained by stacking the emission from the SO 6$_5$-5$_4$ and SO 6$_7$-5$_6$
lines which have similar intensities and excitation conditions. Results of a Gaussian line fit (including opacity correction and hyperfine structure for C$^{17}$O) are indicated in the upper left corner of each panel.
\label{fig:kepler}}
\end{figure}

Figure \ref{fig:kepler} present here the Keplerian corrected profiles for C$^{17}$O 2-1 (top plot) and the SO lines. 
Since this process assumes rotational symmetry, it is appropriate for the SO lines, but non optimal for C$^{17}$O because it ignores the strong azimuthal dependency observed in Fig.\ref{fig:areas}.
The middle plot corresponds to SO $5_6-4_5$ transition, while the bottom plot (SO 6-5) has been obtained by averaging the  SO $6_5-5_4$ and SO $6_7-5_6$ emissions. 

\section{C$^{17}$O fit}
\label{app:co}
\begin{figure}[!ht]
\centering
\includegraphics[width=0.8\columnwidth]{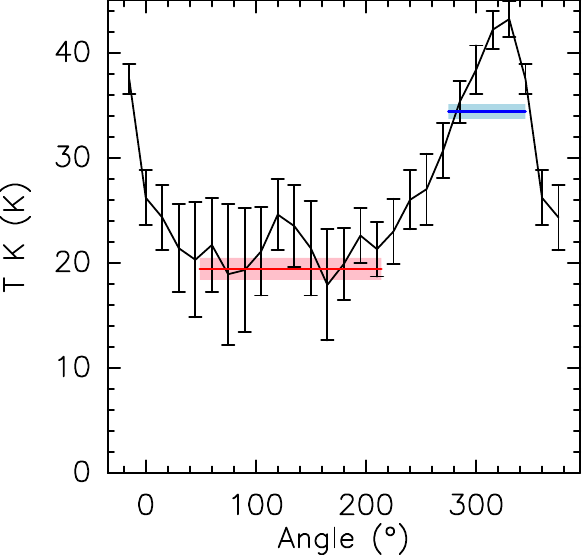}\\
\caption{C$^{17}$O fit of the temperature along the azimuth. 
\label{fig:tk-C17O}}
\end{figure}

Using DiskFit, we derive inner and outer radii of 125 and 286 au (confirming that the C$^{17}$O emission mostly resides inside the SO emission). The limited radial range, combined with the
azimuthal asymmetry, does not allow to constrain the temperature exponent. 
Assuming $q=0.4$, we obtain a reference column density of $4.2\pm0.3\,10^{15}$ cm$^{-2}$ and an exponent $p=2.00\pm0.04$, a temperature of $28\pm4$ K (at 200 au). The distribution being very asymmetric, the errorbars should be taken with caution. 

To quantitatively evaluate the impact of the asymmetry, we extracted the spectra at a radius of 170 au as a function of azimuth. Each spectrum
was fitted using the hyperfine structure of the 2-1 line.
The apparent brightness is given by 
$$T_b(v) = \eta \left(1-\exp(-\tau(v))\right) \left(J_\nu(T_\mathrm{ex}) - J_\nu(T_\mathrm{bg})\right) $$
where $\eta$ is the beam filling factor, and 
the opacity $\tau(v)$ is velocity dependent because of the hyperfine splitting
$$ \tau(v) = \tau \sum_i  x_i \exp\left(-((v-v_i)/\delta v)^2 \right)$$
where $v_i$ is the velocity offset of component $i$ and $x_i$ its relative
intensity, the fit allows to derive both the excitation temperature and the opacity $\tau$ 
of the line
\footnote{The fit residuals show that the hyperfine structure parameters taken from the CDMS database are not optimal. Shifting the lowest velocity component by $+0.13$\,\kms\ improves the agreement with the data, but does not significantly affect the derived values.}.
Furthermore, the critical density of CO isotopologues being small,
$T_\mathrm{ex} \approx T_K$.

Figure \ref{fig:tk-C17O} shows our results, using $\eta=1$. The red line and light-red bar indicate the value
derived by averaging over the $0-180^\circ$ azimuth range to avoid the bias due to the non linearity of the fit, <$T_K$>$=19.7\pm1.0$,\,K. The corresponding blue line shows the average over the $240-300^\circ$ azimuth range. We note that these values are obtained using a linear resolution of $100 \times 55$ au, so that a small (upward) correction to
retrieve the intrinsic temperature should be applied given the relatively narrow ring radial extent. This is different from the DiskFit results that directly refer to the
intrinsic disk model values.

We thus conclude that in the ring, the gas has a temperature above 20 K, peaking above 40 K at the dust trap location. 
\end{appendix}
\end{document}